\PassOptionsToPackage{unicode}{hyperref}
\PassOptionsToPackage{hyphens}{url}
\PassOptionsToPackage{dvipsnames,svgnames,x11names}{xcolor}
\documentclass[
]{article}
\usepackage{amsmath,amssymb}
\usepackage{lmodern}
\usepackage{iftex}
\ifPDFTeX
  \usepackage[T1]{fontenc}
  \usepackage[utf8]{inputenc}
  \usepackage{textcomp} 
\else 
  \usepackage{unicode-math}
  \defaultfontfeatures{Scale=MatchLowercase}
  \defaultfontfeatures[\rmfamily]{Ligatures=TeX,Scale=1}
\fi
\IfFileExists{upquote.sty}{\usepackage{upquote}}{}
\IfFileExists{microtype.sty}{
  \usepackage[]{microtype}
  \UseMicrotypeSet[protrusion]{basicmath} 
}{}
\makeatletter
\@ifundefined{KOMAClassName}{
  \IfFileExists{parskip.sty}{%
    \usepackage{parskip}
  }{
    \setlength{\parindent}{0pt}
    \setlength{\parskip}{6pt plus 2pt minus 1pt}}
}{
  \KOMAoptions{parskip=half}}
\makeatother
\usepackage{xcolor}
\usepackage{color}
\usepackage{fancyvrb}

\DefineVerbatimEnvironment{Highlighting}{Verbatim}{commandchars=\\\{\}}
\newenvironment{Shaded}{}{}

\newcommand{\BuiltInTok}[1]{\textcolor[rgb]{0.00,0.50,0.00}{#1}}

\newcommand{\CommentTok}[1]{\textcolor[rgb]{0.38,0.63,0.69}{\textit{#1}}}

\newcommand{\ConstantTok}[1]{\textcolor[rgb]{0.53,0.00,0.00}{#1}}

\newcommand{\FloatTok}[1]{\textcolor[rgb]{0.25,0.63,0.44}{#1}}
\newcommand{\FunctionTok}[1]{\textcolor[rgb]{0.02,0.16,0.49}{#1}}
\newcommand{\ImportTok}[1]{\textcolor[rgb]{0.00,0.50,0.00}{\textbf{#1}}}

\newcommand{\KeywordTok}[1]{\textcolor[rgb]{0.00,0.44,0.13}{\textbf{#1}}}
\newcommand{\NormalTok}[1]{#1}
\newcommand{\OperatorTok}[1]{\textcolor[rgb]{0.40,0.40,0.40}{#1}}

\providecommand{\tightlist}{%
  \setlength{\itemsep}{0pt}\setlength{\parskip}{0pt}}
\NewDocumentCommand\citeproctext{}{}
\NewDocumentCommand\citeproc{mm}{%
  \begingroup\def\citeproctext{#2}\cite{#1}\endgroup}
\makeatletter
 \let\@cite@ofmt\@firstofone
 \def\@biblabel#1{}
 \def\@cite#1#2{{#1\if@tempswa , #2\fi}}
\makeatother
\newlength{\cslhangindent}
\newlength{\csllabelwidth}
\newenvironment{CSLReferences}[2] 
 {\begin{list}{}{%
  \setlength{\itemindent}{0pt}
  \setlength{\leftmargin}{0pt}
  \setlength{\parsep}{0pt}
  \ifodd #1
   \setlength{\leftmargin}{\cslhangindent}
   \setlength{\itemindent}{-1\cslhangindent}
  \fi
  \setlength{\itemsep}{#2\baselineskip}}}
 {\end{list}}
\usepackage{calc}

\ifLuaTeX
\usepackage[bidi=basic]{babel}
\else
\usepackage[bidi=default]{babel}
\fi
\babelprovide[main,import]{american}

\def\languageshorthands#1{}
\ifLuaTeX
  \usepackage{selnolig}  
\fi
\IfFileExists{bookmark.sty}{\usepackage{bookmark}}{\usepackage{hyperref}}
\IfFileExists{xurl.sty}{\usepackage{xurl}}{} 
\hypersetup{
  pdftitle={Piecewise: Flexible piecewise functions for fast integral
transforms in Julia},
  pdfauthor={Christophe Berthod},
  pdflang={en-US},
  colorlinks=true,
  linkcolor={Maroon},
  filecolor={Maroon},
  citecolor={Blue},
  urlcolor={Blue},
  pdfcreator={LaTeX via pandoc}}

\title{Piecewise: Flexible piecewise functions for fast integral
transforms in Julia}

\definecolor{c53baa1}{RGB}{83,186,161}
\definecolor{c202826}{RGB}{32,40,38}

\usepackage[affil-it]{authblk}
\usepackage{orcidlink}
\author[1%
  ]{Christophe Berthod%
    \,\orcidlink{0000-0002-0787-008X}\,%
    }

\affil[1]{Department of Quantum Matter Physics, University of Geneva,
1211 Geneva, Switzerland%
  }
\date{22 September 2025}

\begin{document}
\maketitle

\section{Summary}\label{summary}

A piecewise function of a real variable \(x\) returns a value computed
from a rule that can be different in each interval of the values of
\(x\). The Julia (\citeproc{ref-Bezanson-2017}{Bezanson et al., 2017})
module \texttt{Piecewise} provides an implementation of piecewise
functions, where the user is free to choose the rules. A mechanism
allows for fitting a piecewise function made of user-defined formulas to
a real function of a real variable. With appropriately chosen formulas,
various integral transforms of the piecewise function become directly
available without relying on quadratures. The module \texttt{Piecewise}
defines seven formula that enable the fast calculation of the moments of
the piecewise function. The module \texttt{PiecewiseHilbert} supplements
these formula with methods enabling a fast Hilbert transform. The module
\texttt{PiecewiseLorentz} extends some of these formula to enable what
we call a Lorentz transform. This code was written to solve a quantum
physics problem involving several coupled integral equations
(\citeproc{ref-MagnetoTransport.jl}{Berthod, 2025};
\citeproc{ref-Morpurgo-2024}{Morpurgo et al., 2024},
\citeproc{ref-Morpurgo-2025}{2025}).

\section{Statement of need}\label{statement-of-need}

The interpolation problem, which consists in constructing a continuous
function out of discrete data, is ubiquitous in many areas of science
and technology. This problem has been traditionally solved by means of
global or piecewise polynomial functions
(\citeproc{ref-Bhagavan-2024}{Bhagavan et al., 2024};
\citeproc{ref-Interpolations.jl}{Kittisopikul et al., 2023};
\citeproc{ref-Numerical-Methods-2021}{Wikibooks, 2021}). The various
interpolation schemes differ by the order of the polynomials, the choice
of the sampling points when this choice is possible, and the additional
conditions required when the solution is not uniquely determined by the
data. Beside drawing a smooth curve through points in a graph, one
important use of interpolations is for constructing a cheap but accurate
approximation of a computer-intensive function. Powerful tools have been
developed for smooth functions that are well approximated by polynomial
interpolants using Chebyshev points (\citeproc{ref-Chebfun}{Driscoll et
al., 2014}; \citeproc{ref-ApproxFun.jl-2014}{Olver \& Townsend, 2014}).
However, if the function presents critical points like discontinuities
or singularities, all polynomial interpolations fail in the neighborhood
of these points, due to the absence of a convergent Taylor series. When
the underlying function has critical points and accuracy is an issue,
there is a need for piecewise interpolation schemes that are based on
nonanalytic functions rather than polynomials.

The mathematical problems involving integral equations (i.e., when the
unknown function appears inside an integral) are often solved
numerically by discretizing the integral and setting up an iteration.
This introduces a discretization error. A choice of the discrete grid
that minimizes the error would generally be nonuniform and require
\emph{a priori} knowledge of the solution. An optimization of the grid
is possible through iterative refinement. However, if the actual
solution has critical points, the iterative refinement will likely fail.
Another approach is to represent the solution at iteration \(n\) as a
piecewise function, evaluate the integrals using quadratures, and fit a
new piecewise function to the solution computed at iteration \(n+1\).
This algorithm may not be faster than the discretization approach, but
it eliminates the discretization bias. Furthermore, the critical points
can in principle be captured in a piecewise function involving
appropriate nonanalytic functions. The procedure requires recursively
fitting a set of elementary functions, including nonanalytic ones that
are problem dependent, to a given function, until a sufficient accuracy
is achieved in each piece. To our knowledge, no Julia package offers
this functionality.

A subclass of all integral equations comprises those involving linear
integral transforms of the kind
\((K\circ f)(\mathbf{X})=\int_{-\infty}^{\infty}dx\,f(x)K(x,\mathbf{X})\),
where \(f(x)\) is a function of a real variable \(x\) and
\(K(x,\mathbf{X})\) is a kernel depending on \(x\) and another, possibly
multidimensional, variable \(\mathbf{X}\). For instance, the \(n\)-th
moment of a distribution is the integral transform of this distribution
with kernel \(K(x,n)=x^n\). Other examples include the Fourier transform
with \(K(x,k)=e^{-ikx}\), the Laplace transform with
\(K(x,s)=\theta(x)e^{-sx}\), \(\theta(x)\) being the Heaviside step
function, the Hilbert transform with \(K(x,y)=1/(y-x+i0^+)\), or more
generally \(K(x,z)=1/(z-x)\) with \(z\in\mathbb{C}\setminus\mathbb{R}\)
a complex number with finite imaginary part. If the function \(f(x)\) is
represented as a piecewise function, and if the various elementary
functions \(F_i(x)\) used in this piecewise representation are such,
that the solution of the differential equation
\(\frac{d}{dx}P_i(x,\mathbf{X})=F_i(x)K(x,\mathbf{X})\) is known
analytically, then \((K\circ f)(\mathbf{X})\) is immediately available
by evaluating the functions \(P_i(x,\mathbf{X})\) at the boundaries of
each piece. This may significantly outperform the evaluation of
\((K\circ f)(\mathbf{X})\) using quadratures, especially near the
critical points of \(f(x)\), where the quadratures typically converge
slowly, if they converge at all. Thus, an environment where
problem-dependent functions \(F_i(x)\) may be defined and used in
piecewise functions, together with kernel-dependent functions
\(P_i(x,\mathbf{X})\), is desirable.

\section{\texorpdfstring{The \texttt{Piecewise}
modules}{The Piecewise modules}}\label{the-piecewise-modules}

The module \texttt{Piecewise} provides such an environment based on
three structures. A structure called \texttt{Formula} holds a
user-defined function depending on a given number of parameters,
together with possible restrictions regarding the values of these
parameters with respect to the interval in which the formula is used. A
second structure called \texttt{Piece} holds an interval, a rule that
can be a sum of \texttt{Formula} objects, and the parameters to be
passed to these formula. Finally, a structure called
\texttt{PiecewiseFunction} holds a collection of \texttt{Piece} objects.
The module comes with a small set of pre-defined \texttt{Formula} that
should cover a wide variety of cases and a method for fitting a
\texttt{PiecewiseFunction} object to a given function. The pre-defined
\texttt{Formula} earn additional methods in the module
\texttt{PiecewiseHilbert}, such that the Hilbert transform of piecewise
functions using these formula can be evaluated without using
quadratures. The module \texttt{PiecewiseLorentz} offers this
functionality for another integral transform (see the
\href{https://christopheberthod.github.io/Piecewise.jl/dev/lorentz.html}{documentation}).

\section{Examples}\label{examples}

\begin{enumerate}
\def\labelenumi{\arabic{enumi}.}
\tightlist
\item
  Two tutorials are available as Jupyter notebooks:
\end{enumerate}

\begin{itemize}
\tightlist
\item
  \href{https://github.com/ChristopheBerthod/Piecewise.jl/blob/main/notebooks/Tutorial-1.ipynb}{Tutorial
  1}: Constructing approximations with
  \href{https://christopheberthod.github.io/Piecewise.jl/dev/index.html\#Piecewise.piecewisefit}{\texttt{Piecewise.piecewisefit}}
\item
  \href{https://github.com/ChristopheBerthod/Piecewise.jl/blob/main/notebooks/Tutorial-2.ipynb}{Tutorial
  2}: Solving an implicit equation using
  \href{https://christopheberthod.github.io/Piecewise.jl/dev/hilbert.html}{\texttt{PiecewiseHilbert}}
\end{itemize}

\begin{enumerate}
\def\labelenumi{\arabic{enumi}.}
\setcounter{enumi}{1}
\item
  For a complete use case, see
  \href{https://github.com/ChristopheBerthod/MagnetoTransport.jl}{\texttt{MagnetoTransport.jl}}.
\item
  For an example of nonlinear integral equation solved using
  \texttt{Piecewise}, see van der Marel \& Berthod
  (\citeproc{ref-vanderMarel-2024}{2024}).
\item
  A simple demonstration is provided below. It is an abbreviated form of
  \href{https://github.com/ChristopheBerthod/Piecewise.jl/blob/main/notebooks/Tutorial-1.ipynb}{Tutorial
  1}, where ample explanation and details are given.
\end{enumerate}

Electronic density of states (DOS) functions typically have critical
points. The DOS is derived from a dispersion relation
\(\varepsilon(\mathbf{k})\) as
\(N(E)=\int\frac{\mathrm{d}^dk}{(2\pi)^d}\delta\big(E-\varepsilon(\mathbf{k})\big)\)
in dimension \(d\), where \(\delta(\cdot)\) is the Dirac delta function.
\(N(E)\) has critical points whenever
\(\nabla\varepsilon(\mathbf{k})=0\) for some \(\mathbf{k}\). The DOS is
an ingredient of many calculations, but in general it is not known
analytically. In the following illustration, we construct a one-piece
approximation with relative accuracy below \(10^{-5}\) for such a DOS
function.

Electrons hopping with unit energy between neighboring sites of a
two-dimensional square lattice with unit lattice parameter have a
dispersion relation \(\varepsilon(\mathbf{k})=2(\cos k_x+\cos k_y)\).
This case is peculiar in that the DOS is known analytically:
\(N(E)=K\big(1-(E/4)^2\big)\theta(4-|E|)/(2\pi^2)\) with \(K\) the
elliptic function. It has two discontinuities at \(E=\pm4\) and a
logarithmic singularity at \(E=0\).

We proceed as follows to construct a piecewise approximation to the DOS.

\begin{enumerate}
\def\labelenumi{\arabic{enumi}.}
\tightlist
\item
  We define the DOS (here: an explicit formula; in general: the result
  of a numerical quadrature).
\item
  We remove the known logarithmic singularity --- represented as a
  \texttt{PiecewiseFunction} --- from the DOS.
\item
  We fit a polynomial to the residual (the algorithm automatically
  chooses order 9 to achieve the requested accuracy) and we add the
  logarithmic singularity back.
\item
  We check the error of the approximation.
\end{enumerate}

\begin{Shaded}
\begin{Highlighting}[]
\NormalTok{julia}\OperatorTok{\textgreater{}} \ImportTok{using} \BuiltInTok{SpecialFunctions}\NormalTok{: ellipk}
\NormalTok{julia}\OperatorTok{\textgreater{}} \ImportTok{using} \BuiltInTok{Piecewise}

\NormalTok{julia}\OperatorTok{\textgreater{}} \CommentTok{\# DOS function. Due to the dependence on E\^{}2, ellipk(1 {-} (E / 4)\^{}2)}
\NormalTok{julia}\OperatorTok{\textgreater{}} \CommentTok{\# loses accuracy for |E| \textless{} 1e{-}4. We use the known expansion instead.}
\NormalTok{julia}\OperatorTok{\textgreater{}} \FunctionTok{N}\NormalTok{(E) }\OperatorTok{=}\NormalTok{ (}\FunctionTok{abs}\NormalTok{(E) }\OperatorTok{\textless{}} \FloatTok{1e{-}4}\NormalTok{ ? }\FunctionTok{log}\NormalTok{(}\FloatTok{16} \OperatorTok{/} \FunctionTok{abs}\NormalTok{(E)) }\OperatorTok{:} \FunctionTok{abs}\NormalTok{(E) }\OperatorTok{\textgreater{}} \FloatTok{4}\NormalTok{ ? }\FloatTok{0.0} \OperatorTok{:}
           \FunctionTok{ellipk}\NormalTok{(}\FloatTok{1} \OperatorTok{{-}}\NormalTok{ (E }\OperatorTok{/} \FloatTok{4}\NormalTok{)}\OperatorTok{\^{}}\FloatTok{2}\NormalTok{)) }\OperatorTok{/}\NormalTok{ (}\FloatTok{2} \OperatorTok{*} \ConstantTok{pi}\OperatorTok{\^{}}\FloatTok{2}\NormalTok{);}

\NormalTok{julia}\OperatorTok{\textgreater{}} \CommentTok{\# Piecewise function representing the logarithmic singularity}
\NormalTok{julia}\OperatorTok{\textgreater{}}\NormalTok{ singularity }\OperatorTok{=} \FunctionTok{PiecewiseFunction}\NormalTok{(}\OperatorTok{:}\NormalTok{even,}
           \FunctionTok{Piece}\NormalTok{((}\FloatTok{0}\NormalTok{, }\FloatTok{4}\NormalTok{), (}\ConstantTok{false}\NormalTok{, }\ConstantTok{true}\NormalTok{), LOG, [}\FloatTok{0}\NormalTok{, }\OperatorTok{{-}}\FloatTok{1} \OperatorTok{/}\NormalTok{ (}\FloatTok{2} \OperatorTok{*} \ConstantTok{pi}\OperatorTok{\^{}}\FloatTok{2}\NormalTok{)]));}

\NormalTok{julia}\OperatorTok{\textgreater{}} \CommentTok{\# Remove the singularity before fitting and add it back afterwards}
\NormalTok{julia}\OperatorTok{\textgreater{}} \CommentTok{\# PiecewiseFunction objects can be summed.}
\NormalTok{julia}\OperatorTok{\textgreater{}}\NormalTok{ f }\OperatorTok{=} \FunctionTok{piecewisefit}\NormalTok{(E }\OperatorTok{{-}\textgreater{}} \FunctionTok{N}\NormalTok{(E) }\OperatorTok{{-}} \FunctionTok{singularity}\NormalTok{(E),}
\NormalTok{           (}\FloatTok{0}\NormalTok{, }\FloatTok{4}\NormalTok{), [POLY], parity}\OperatorTok{=:}\NormalTok{even, rtol}\OperatorTok{=}\FloatTok{5e{-}6}\NormalTok{);}
\NormalTok{julia}\OperatorTok{\textgreater{}}\NormalTok{ f }\OperatorTok{+=}\NormalTok{ singularity}
\OperatorTok{\textless{}}\NormalTok{ Piecewise even }\KeywordTok{function}\NormalTok{ with }\FloatTok{1}\NormalTok{ piece and support [}\OperatorTok{{-}}\FloatTok{4.0}\NormalTok{, }\FloatTok{4.0}\NormalTok{] }\OperatorTok{\textgreater{}}

\NormalTok{julia}\OperatorTok{\textgreater{}} \CommentTok{\# Check that the relative error is smaller than 1e{-}5}
\NormalTok{julia}\OperatorTok{\textgreater{}} \FunctionTok{maximum}\NormalTok{(}\FunctionTok{LinRange}\NormalTok{(}\OperatorTok{{-}}\FloatTok{4}\NormalTok{, }\FloatTok{4}\NormalTok{, }\FloatTok{1000}\NormalTok{) }\OperatorTok{.|\textgreater{}}\NormalTok{ E }\OperatorTok{{-}\textgreater{}} \FunctionTok{abs}\NormalTok{(}\FunctionTok{f}\NormalTok{(E) }\OperatorTok{./} \FunctionTok{N}\NormalTok{(E) }\OperatorTok{{-}} \FloatTok{1}\NormalTok{)) }\OperatorTok{\textless{}} \FloatTok{1e{-}5}
\ConstantTok{true}

\NormalTok{julia}\OperatorTok{\textgreater{}} \CommentTok{\# Printing a PiecewiseFunction shows the constructor for that object.}
\NormalTok{julia}\OperatorTok{\textgreater{}} \CommentTok{\# The exact numbers may vary, as randomness is involved in the fitting.}
\NormalTok{julia}\OperatorTok{\textgreater{}} \FunctionTok{println}\NormalTok{(f)}
\FunctionTok{PiecewiseFunction}\NormalTok{(}\OperatorTok{:}\NormalTok{even, [}
    \FunctionTok{Piece}\NormalTok{((}\FloatTok{0.0}\NormalTok{, }\FloatTok{4.0}\NormalTok{), (}\ConstantTok{false}\NormalTok{, }\ConstantTok{true}\NormalTok{), [POLY, LOG],}
\NormalTok{        [[}\FloatTok{1.404609620501190e{-}01}\NormalTok{, }\FloatTok{1.174637035445451e{-}04}\NormalTok{, }\FloatTok{2.462481494536943e{-}03}\NormalTok{,}
        \OperatorTok{{-}}\FloatTok{1.995071066151247e{-}03}\NormalTok{, }\FloatTok{1.349285972889321e{-}03}\NormalTok{, }\OperatorTok{{-}}\FloatTok{6.607932327688245e{-}04}\NormalTok{,}
        \FloatTok{2.158678002039584e{-}04}\NormalTok{, }\OperatorTok{{-}}\FloatTok{4.417944490515085e{-}05}\NormalTok{, }\FloatTok{5.103263826300048e{-}06}\NormalTok{,}
        \OperatorTok{{-}}\FloatTok{2.533300199959029e{-}07}\NormalTok{], [}\FloatTok{0.000000000000000e+00}\NormalTok{,}
        \OperatorTok{{-}}\FloatTok{5.066059182116889e{-}02}\NormalTok{]])}
\NormalTok{])}
\end{Highlighting}
\end{Shaded}

\section*{References}\label{references}
\addcontentsline{toc}{section}{References}

\phantomsection\label{refs}

\end{document}